\documentclass[10pt]{article}
\begin{document}
{\bf MODEL CATEGORIES AND QUANTUM GRAVITY}

\bigskip

by: {\bf LOUIS CRANE; Math Department, KSU}

\bigskip

{\bf ABSTRACT:} We propose a mathematically concrete way of modelling the suggestion that in quantum gravity the spacetime disappears, replacing it with a discrete approximation to the causal path space described as an object in a model category. One of the versions of our models appears as a thickening of spacetime, which we interpret as a formulation of relational geometry. Avenues toward constructing an actual quantum theory of gravity on our models are given a preliminary exploration.

\bigskip

{\bf I. INTRODUCTION}

\bigskip

It has been shown recently \cite{Susskind} that a bounded region in a spacetime in general relativity can only transmit a finite amount of information to the exterior, proportional to its boundary area in Planck units.
If we want to construct a quantum theory of GR which would describe the results of experiments which external observers perform on bounded regions of spacetime, then the description of spacetime as a point set continuum contains too much information. 

It was the view of Einstein \cite{Stachel} that the
excess of information of a metric on a continuum spacetime is the source of the problems at the core of quantum field theory and quantum gravity.

More recently \cite{AB} it has been proposed that the continuum disappears in the quantum theory of general relativity.

There exists a model for quantum gravity, the BC model \cite{BC} which is finite on any finite four dimensional simplicial complex. This however leaves the problem of choosing the particular triangulation of spacetime. On the other hand there is the GFT model \cite{Rovelli}, which sums over all triangulations. Unfortunately, this is an infinite sum.

We have long believed that incorporating the finite information results for a bounded region into a more sophisticated model intermediate between BC and GFT, would be the resolution of this dilemma, but that that would need a more subtle type of topological structure.

In this paper we search for models for the topology of a region of spacetime R which only contain information accessible to an external observer in general relativity. In order to do this, we make a physical conjecture that one type of information, namely the multiple images of a light source exterior to R as viewed by an observer through R, can give a complete description of the observable state of R. We call this the {\bf lensing hypothesis}, and discuss its plausibility below.

Assuming our conjecture, we show that well understood and very computable models for the topology of spacetime regions exist. There are several equivalent constructions, going under the names of Quillen \cite{Quillen} and Sullivan \cite{Sullivan} models. The mathematical reason we can do this is that some of the multiple images of the light source correspond to rational cohomology classes  of the loop space of R via Morse theory. The rest of the images also fit into the Sullivan models in a natural way as contractible algebras. Perhaps the most interesting version of the model, in light of the developments of categorical state sums for TQFTs and quantum general relativity \cite{CY, CF}, is a differential graded Hopf algebra.

Throughout this paper, we will assume that all the regions we study are simply connected. Mathematically, this is because the model category constructions we use only work in that case. The theorems we state would need to be weakened for multiply connected regions. Physically, we feel comfortable with this restriction because it is hard to imagine a multiply connected region on a macroscopic scale which could be treated in isolation from its surroundings. On a microscopic scale, we are effectively ruling out wormholes, which are incompatible with the positive energy condition. It would be possible, although more difficult, to weaken this assumption.

The family of functors Quillen constructed allow us to reconvert the models into cellular and simplicial complexes in two different ways, raising the possibility of a BC type state sum on them.

At the end of the paper, we begin a discussion of how to construct quantum theories over these models which could describe gravity and matter. This subject is still a work in progress, but there are several promising directions.

Before we go on to explain the constructions of these models and how they relate to physical processes, 
we shall say a little about the philosophy which led us to them. In another famous quote, Einstein said that the experience of measuring the surface of the earth, which was the origin of geometry, led to the intuition that spacetime is a kind of substance, which he believed to be profoundly misleading. 

I think the idea of a fixed absolute spacetime point set, independent of the observer, is the result of such an intuition. The classical continuum, which we got from Euclid via Newton and Weierstrauss, is an extension of the results of everyday experience to a domain in which they do not apply. There is no good reason to suppose that a spacetime interval shorter than the Planck scale can be infinitely subdivided, and importing the assumption that it can via the trojan horse of background coordinates leads quantum theories of gravity into the swamp of the ultraviolet divergences.

A very important motivation for this paper is the distinction between absolute and relational position. A region in spacetime thought of as a point set has its ``position'' determined by what set of points it is. This idea of position is unphysical. Rather, the foundation of a quantum theory of gravity should be {\it relational} position, by which I understand where some region appears to be to all external observers \cite{newcrane}. 

Now if we imagine two different spacetime metrics on a given bounded region, a subregion in one will have a different relational position from any subregion of the other. This is because the curvature of the metrics will bend light rays going in different directions by different amounts, so that a subregion that appeared to be in the same place as a subregion in a different metric to one outside observer would not appear so if observed by another observer from a different direction.

Thus a region in a quantum state for the spacetime metric would have more subregions than a classical region. They would be arranged in layers, so that the quantum region would appear like a thickening of the classical region.

I think of this as like a sheet of paper. At length scales large compared to its thickness, a sheet of paper can be described as a smooth surface, but at shorter scales it is a very complex web.

If we are serious about abandoning the absolute point set background, we need to take the point of view that quantum gravity lives on some such thickening.

For this reason, I think it is interesting that one of the collection of functors we will be studying, namely the geometric realisation of Sullivan algebras, produces precisely such a thickening. We will explore near the end of this paper the possibility of constructing a state sum model for general relativity on the geometric realisation of a Sullivan algebra. 

A region of spacetime which is to be treated as quantum cannot have classical observers inside it. Like any quantum process it can only be discussed after it is complete. We must not ask what it looks like on the inside, but only about regularities of relationships between external observations.  

Like any quantum system, GR must have observables that are complementary, that is cannot simultaneously possess sharp values. Already the development of the BC model \cite{BC} shows us that different geometrical data such as areas of faces cannot have simultaneous sharp values.

This has implications for the topology appropriate to the descriptions of spacetime regions. Rather than a single topology, we should expect to find complementary descriptions of the topology appropriate to complementary ways of observing the geometry. This may seem hard to imagine within the framework of point set topology, but modern homotopy theory has many structures which can be considered to have a topology, and natural approaches exist connecting them to different ways of probing the spacetime geometry of a region.

The different forms of the Quillen and Sullivan models we shall describe below suggest complementary descriptions of spacetime, as we shall indicate. The reason is that some of the models correspond to discretizations of the spacetime region itself, and some to discretizations of its loop space, or causal path space.

\bigskip

{\bf II. SOME THOUGHTS ON THE LENSING HYPOTHESIS}

\bigskip

In a sense the lensing hypothesis is rather natural. If a region of spacetime is too small to enter by a classical observer, there isn't much else to do except "hold it up to the light". A classical observer inside a region of spacetime would decohere it, so unless we want to think about many worlds pictures, it is contradictory for a quantum region to have an internal observer.

Also, since as we mentioned above, a bounded region can only transmit a finite amount of information to its exterior, and since most nontrivial topologies have an infinite number of images in a classical spacetime description of lensing, \cite{Perlick} there is certainly enough data.

What we directly see when we observe a curved region of spacetime is only a discrete decomposition of its space of causal paths. Mathematicians have discovered that such decompositions have algebraic structure (inherited from the multiplication on loop space) which are handy descriptions of topological spaces. Such are the Quillen models which we are proposing to use in GR.

A natural objection to the lensing hypothesis is that a large flat region of spacetime transmits only a single image of a light source. I believe the resolution of this is that a classical state of any quantum system emerges as in the decoherence picture as  a superposition of many similar quantum states. 

We can then invoke the suggestion of Rovelli \cite{Rovelli2} that a particle travelling through a quantized spacetime would automatically appear quantized, since even classical GR imposes the equations of motion on any particle travelling through it. Combining the idea of many similar quantum states reinforcing with Rovelli's suggestion, we end up with the picture that the single fuzzy quantum image of a source is the superposition of many multiple images in which the source is scattered by many smaller strongly curved quantum regions, obtaining the quantum propagator from a kind of random walk.

In other words, the quantum mechanical uncertainty of the positions of optical images can be explained as a superposition of the multiple images of an ensemble of spacetime geometries, each highly curved near the Planck scale.

If this picture could be precisely implemented, it would add strength to the lensing hypothesis.

Finally, to be honest, we must invoke the streetlight principle. The mathematical picture which results from the hypothesis is rich and computationally accessible, so we investigate it because we can.

It would be possible also to state a {\bf weak lensing hypothesis,} namely that the locations and intensities of the multiple images define an interesting sector of quantum gravity in a region. We choose not to preface every result in the following with the phrase "in the lensing sector," but the reader may do so. 

\bigskip

{\bf III. DESIDERATA FOR A DESCRIPTION OF QUANTUM REGIONS}

\bigskip
If a region of spacetime is not described mathematically by a point set, then some other mathematical structure is necessary.

Now let us list the properties a mathematical description of a region of quantum spacetime should have. We shall include the lensing hypothesis in our assumptions. 
\bigskip

{\it A model of quantum spacetime should:}

\bigskip

{\it A. Assign a mathematical structure Q(R) to each region of spacetime.}

\bigskip

{\it B. Assign to each mapping of regions $L : R  \longrightarrow S$ a map of Q structures }

\bigskip
{\it $ Q(L) : Q(R) \longrightarrow Q(S)$ , }

\bigskip

{\it which takes compositions to compositions.}

\bigskip

{\it C. Have a class of homotopies on maps of Q structures so that given any two maps of one region to another which are homotopic (can be smoothly deformed into one another), there exists a homotopy of the corresponding maps on Q structures, again preserving composition.}
\bigskip

{\it D. The structure Q(R) should have enough information to compute the probability amplitudes for the apparent positions and arrival times of the multiple images of light sources seen through R, and be reconstructable from this information.}

\bigskip

The physical motivation for this list is that subregions can be effectively included, and the principle of diffeomorphism invariance, so central to general relativity, can be expressed; together of course, with the lensing hypothesis.

The mathematical motivation is that the construction be functorial or actually 2-functorial. That is the operation Q is actually a functor from the 2-category of regions, smooth maps and smooth homotopies to a category of models for quantum regions, also equipped with maps and homotopies. 

The central purpose of the rest of this paper is to show that the problem of constructing models satisfying our list is already solved in the context of rational homotopy theory by the theory of model categories. This is due to a string of mathematical coincidences which is actually rather obvious once one has assembled the necessary background.

\bigskip

{\bf IV. MORSE THEORY, GEODESICS, AND LENSING}

\bigskip

Morse theory is a branch of smooth topology which obtains information about the topology of a smooth manifold by studying the singularities of generic smooth functions on it. A function is generic if it has only singularities of the lowest order, i.e. if the first derivatives only all vanish at isolated points, and the matrix of second derivatives at each such point is nonsingular. For a good introduction, see \cite{Milnor}. 

When one has a Morse function on a manifold, the gradient flow down to the singularities divides the manifold into standard structures called handles. Thus a Morse function is a snapshot of a handlebody decomposition of the manifold.

Critically important for us are the Morse inequalities. These tell us that a Morse function on a manifold M must have at least as many singularities of degree d as the number of generators of the cohomology of M in dimension d with real coefficients. A perfect Morse function has just so many singularities; in general a Morse function may have more than these, but they come in cancelling pairs of adjacent dimensions.
The reason we know this is a branch of mathematics called Cerf theory \cite{Cerf}, which studies how we can move from one Morse function to another, moving through intermediate functions with only the simplest types of higher singularities. This gets us from one Morse function to another by a series of ``moves,'' and the only move which changes the number of singularities adds two cancelling handles in adjacent dimensions, not changing the topology.

The singularities of Morse functions only correspond to the cohomology with real (or equivalently rational) coefficients; cocycles which are torsion (some multiple is zero) play no part. Cohomology with real coefficients is always a real vector space with the same dimension as the rational dimension of the rational cohomology. Mathematicians like to speak of rational cohomology, or sometimes cohomology with coefficients in an arbitrary field of characteristic zero. Physicists prefer to speak of real cohomology. I shall speak of them interchangeably.

The classical Morse theory was for finite dimensional manifolds, but infinite dimensional Morse theory has been developed by many authors [11, 12, 13]. An important application was the study of closed geodesics on Riemannian manifolds. If we consider the space of smooth closed loops on M with a given basepoint $ \Omega (M)$, which is a smooth infinite dimensional manifold, the length, or better the energy (see \cite{Milnor}) of the path turns out to be a Morse function generically. Thus the problem of closed geodesics on M is related to finding the rational cohomology of $ \Omega (M)$. 

Now the space of paths in M with fixed endpoints p and q is very similar to the space of closed loops. This is because the smooth path space P(p,q,M) is of the same homotopy type as $ \Omega(M)$. Hence counting the geodesics with fixed endpoints is a very similar problem to the classical one of counting closed geodesics. This is another result from topology which will be important to us.

Now let us turn to the problem of gravitational lensing, or describing the multiple images which an observer in a spacetime may see of a light source in its causal past.

Fermat's principle continues to hold in general relativity. Classical light rays correspond to null geodesics in the spacetime. Thus the problem of counting and describing the images an observer sees of a pointlike light source is closely analogous to finding the geodesics in a Riemannian manifold with fixed endpoints, but with a Lorentzian signature metric.

The spacetime approach to gravitational lensing rests on an adaptation of Morse theory to the spacetime case \cite{Perlick, Uhlenbeck}. One has to modify the application of Morse theory by considering the space of all smooth null paths which begin at a fixed event x in a spacetime S and end at some point on the worldline of an observer y(t). This is because different null geodesics have different arrival times. We denote this infinite dimensional space P(x,y,S).

The arrival time acts as the Morse function. Once again, it is closely related to an infinite dimensional space, which in physically reasonable cases (i.e. globally hyperbolic spacetimes) is of the same homotopy type as the loop space of the spacetime \cite{Perlick}. 

If the time lapse happens to be a perfect Morse function, then the images will count a set of generators of the cohomology ring of the loop space $H_Q ( \Omega (S))$ for the reasons we have discussed. In general, this is not so. There can also be other images which come in pairs, with opposite orientations. These pairs of images disappear by meeting at "folds in the sky" as the position of the observer or source is varied. The folds in the sky appear as caustics in lensing images, and have been photographed by astronomers. Folds in the sky correspond to cancelling handles.

\bigskip

{\bf EXAMPLE: SCHWARTZSCHILD BLACK HOLE}

\bigskip

It is known from an exact calculation \cite{K} that the Schwartzschild black hole has an infinite series of images as a gravitational lens. This reflects the fact that orbits in general relativity are not closed, and can circle the central mass any number of times before escaping to infinity.

It would be a difficult problem to determine directly what any perturbation of the Schwartzschild solution might do to its lensing properties. It is therefore of interest that the infinite number of images can be predicted by a purely topological calculation \cite {Perlick}.

This is possible because the Schwartzschild solution has the homotopy type of $S^2$. Thus the number of lensing images is at least the dimension of $H_Q ( \Omega (S^2))$, which is known to be infinite.

 As we shall see below, calculation of the rational cohomology of the loop groups of spheres was one of the seminal problems for rational homotopy theory, and can most easily be solved by means of model category theory.

At this point we can already see that the lensing hypothesis leads us into a situation where the information we would observe about a region is closely related to rational cohomology of loop spaces and therefore to rational homotopy theory.

\bigskip

{\bf V. RATIONAL HOMOTOPY THEORY, MODEL CATEGORIES AND LOOP SPACES. }

\bigskip

If we were able to observe the lensing effect of a highly curved region of spacetime, the images that did not disappear in pairs as we rotated our position of observation around the region would correspond to a set of generators for the rational cohomology of the region. (It is also possible for a cancelling pair to appear and for one of the new handles to cancel an image formerly representing a cohomology cycle, leaving its partner to represent the cycle, but that doesnt change the point). This suggests that a quantum theory of the lensing sector of gravity could be constructed by replacing the region with its rational homotopy type. It will be seen that the  images which cancel in pairs and therefore represent cancelling handles also fit into the models for rational homotopy theory as contractible models.

We remind the reader that in this paper all regions are assumed to be simply connected.

In fact the Quillen and Sullivan models are sophisticated descriptions of discrete decompositions of spacetime or its causal path space. They inherit their algebraic structure from the multiplication on the loop space or the multiplication of differential forms.

The rational homotopy type of a space is much simpler than its general homotopy type. The rational homotopy and cohomology groups of simple spaces can be computed exactly, whereas the homotopy and cohomology groups including the torsion, are not all simultaneously known for any nontrivial space, one being hard whenever the other is easy.

The calculations of rational homotopy and cohomology proceed by replacing the space with a model for its rational homotopy type. The calculations for loop spaces proceed by finding a model loop space for the model. There are several different constructions, which have been shown to be equivalent \cite{M}. In this section we give a nontechnical introduction to these mathematical ideas.

\bigskip
{\bf A. RATIONAL HOMOTOPY TYPE}

\bigskip

Two maps $ L_1, L_2 : A \rightarrow B$ are homotopic if there exists a map $ L: A \times I \rightarrow B $ such that
$L(a,0) = L_1 (a), L(a,1) = L_2 (a) $. L is called a homotopy.

Two spaces A and B are said to have the same homotopy type if there are continuous maps $f: A \rightarrow B$ and $ g: B \rightarrow A $ such that both $ f \circ g$ and $g \circ f$ are  homotopic to the identity. For most purposes, we can think of A and B as topologically the same in this case.  A weak homotopy equivalence is a map between two spaces which induces an isomorphism on cohomology. In many cases this implies homotopy equivalence.

Now we can define a rational homotopy equivalence as a map between two spaces which induces an isomorphism on the rational cohomology groups of the two spaces. The equivalence class of a space under this equivalence relation is its rational homotopy type. 

Note that the existence of a map is necessary in this definition. Two spaces whose rational cohomology groups are isomorphic as groups need not be rationally homotopy equivalent. If two spaces are rational homotopy equivalent, it does follow that their rational cohomology is isomorphic.

Any space has a rationalisation, i.e. a map to a space whose homotopy and cohomology groups are rational (that is, torsion free), such that the map is a rational homotopy equivalence.

For example, none of the spheres are rational, since they all have torsion in their higher homotopy groups. Even the complete homotopy groups of $S^2$ are unknown. By contrast the rational part of the homotopy groups of $S^2$ are generated by one generator in dimension 2 and one in dimension 3. 

The reason rational homotopy theory is so computable is that the rational homotopy category is equivalent to a number of model categories. Put more concretely, a space can be replaced, for the purposes of rational homotopy theory, by a mathematical structure which is simpler and can be computed with and constructed on as if it were geometrical. This can be done several ways; and the constructions have strong resonances with mathematical Physics.

\bigskip

{\bf B. THE QUILLEN AND SULLIVAN MODELS}

\bigskip
The first models for rational homotopy types appear in Quillen \cite{Quillen} who had already formalized the notion of a model category in \cite {Quillen2}. (I reccommend that anybody not primarily grounded in homotopy theory learn about the examples first and the formalisation only as needed.)

Quillen proved that the category of rational homotopy types of topological spaces is equivalent to several other categories. One is the category of complete cocommutative differential graded Hopf algebras (A graded algebra or coalgebra is graded (co)commutative if two odd indexed elements anti(co)commute, and any other combination (co)commutes).   

Quillen's construction proceeds by a series of transformations of the topological space. Each transformation is shown to be a homotopy equivalence of categories. Each category in the chain is an abstract model category, which means that objects in it can be thought of as having topology, or differently put, that homotopy theory can be done inside the category.

Each transformation is expressed by a pair of adjoint functors that go both ways between the two categories. This means that each transformation can be reversed, without changing the rational homotopy type. This possibility of going back and forth between different constructions is what leads me to believe that the Quillen mathematical setting is a flexible and powerful one. In particular, it is possible to change the algebraic models back into cellular or simplicial complexes, thus opening the possibility of putting categorical state sum models for gravity, such as the BC model, on them.

In the interest of accessibility of exposition, we shall not discuss the formal definition of a model category; this can be found in \cite{Quillen2}, or a good number of more recent books.

Quillen originally did only the simply connected case, this can be generalized, but will be sufficient for our purposes.

The first step is begun by replacing the space X with a simplicial complex. A physicist might think of this as triangulating the space, then describing the space only by the discrete combinatorial structure of the triangulation. Quillen actually replaces the space X with the complex C(X) of all singular simplices in X. (A singular simplex in X is a map of the standard simplex into X.) This is homotopy equivalent to the naive picture, but more flexible. As a downside, it produces an enormous space, which is difficult to compute with directly.

The first step is concluded by replacing the simplicial complex by a 2-reduced simplicial complex, i.e. one in which there is only one vertex. This can always be done for a simply connected complex by contracting a maximal tree in its 1-skeleton. This last is not strictly necessary, but greatly simplifies the subsequent algebra.

In the second step, the 2-reduced complex C(X) is replaced by the free group (group of strings) of simplices in C(X). This is denoted $C( \Omega (X))$. The reason for the notation is that $C( \Omega (X))$ is a complex for the loop space of X. (Think of a string of simplices as the family of paths in X which traverse those simplices in that order.) Strings are multiplied by concatenation, the inverse of a string traverses the simplices backwards and in the reverse order.

$C( \Omega (X))$ is an example of a simplicial group, which is a simplicial complex where the simplices of each dimension form a group and the boundary and degeneracy operations are homomorphisms. The singular simplicial complex C(G) of a topological group G is a simplicial group under pointwise multiplication of simplices, and the simplicial group structure of $C( \Omega (X))$ comes from the natural group structure on $ \Omega (X)$. Quillen also proves that any reduced simplicial group comes from a space in this way.

The third step is to replace the simplicial group with the differential graded Hopf algebra $GHA( \Omega (X))$ of rational functions on the set of its simplices, then completing it in a certain topology. A differential graded Hopf algebra is a differential graded vector space, (i.e. a sequence of spaces $V_i$ for each integer i with a map $d_i: V_i \rightarrow V_{i-1} $  with $d^2=0$) with a multiplication and comultiplication which are additive in degree and satisfy the axioms of a Hopf algebra. The multiplication in this Hopf algebra comes from composition in $\Omega(X)$; its comultiplication is the standard comultiplication on simplices, dual to the cup product of cohomology. In this section the definition of Hopf algebra we are using includes the assumption that it is graded cocommutative. In the section on the Sullivan model, we are in a dual picture, so Hopf algebras are graded commutative.

Since this is in particular a category of graded vector spaces, it is no surprise that the torsion part of the homology has disappeared, so the ordinary homotopy theory of complete DG Hopf algebras corresponds to the rational homotopy theory of the original space. The definition of the intrinsic homotopy theory of this category is quite technical.

As in the earlier steps, Quillen actually shows that the category of ``complete'' (I am suppressing that part of the definition for brevity) cocommutative DG-Hopf Algebras is homotopy equivalent to the earlier categories, in particular the category of rational homotopy types of simply connected spaces.

After this, Quillen goes on to show that the class of Hopf algebras we are interested in are all universal enveloping algebras of differential graded Lie algebras, so that there are also differential graded Lie algebra models. The Lie algebras can be recovered as the primitive elements of the Hopf algebras, i.e. those which are not non-trivial products.

The Lie algebra model for X has the property that its homology is a graded Lie algebra, which corresponds to the rational homotopy groups of X with the Whitehead product \cite{Book} for bracket. The rational homology of the loop space of X is the universal enveloping algebra of the Lie algebra of the rational homotopy groups of X
 
We will not need here to examine Quillen's last model, in which he replaces the DG Lie algebra by the DG coalgebra of its bar construction, or classifying space.

Another transformation is possible which Quillen did not develop. We can replace the DG Hopf algebra by the category of its representations 
$Rep(HA( \Omega (X)))$. This will be worked out in detail in a subsequent paper. Given the analogous roles played by Hopf algebras and tensor categories of representations in the construction of TQFTs \cite{CY, CF}, this further transformation is very natural for our purposes. We will comment on the physical interest of this new variant of Quillen's model below.

Sullivan \cite{Sullivan} discovered a new approach to constructing models for rational homotopy theory. He began by constructing the graded algebra of forms   on a simplicial complex. To each simplex S he associated the differential graded commutative algebra of differential forms $ \Lambda (S) $ whose coefficients were polynomial in the coordinates on the simplex. (A  singular n-simplex in X is a map of the standard n-simplex in $R^{n+1} $ into X , so it inherits natural coordinates). He then made the natural boundary restrictions on forms, and  defined the differential graded algebra on the simplicial complex X as the subset of the product over simplices S of X of   $ \Lambda (S) $ where forms on each S agree with their restrictions to the boundary of S. The resulting graded commutative algebra is denoted A(X).

In simpler words, A(X) is the differential graded algebra of forms on X which are polynomial in all the simplices of some  triangulation of X.

If X is a smooth manifold, then the resulting differential graded algebra is homotopy equivalent to the de Rham complex of smooth differential forms on X.

Now since Sullivan has started with forms, which are contravariant, the new model is dual as a graded vector space to the Quillen models. If we ask for a complete commutative DG Hopf algebra which is homotopy equivalent to a Sullivan model, its Lie algebra of pure elements is dual to the rational homotopy groups of X.

The Sullivan models admit transformations back to simplicial complexes by means of a functor called geometrical realization. It is constructed by assigning n-simplices to all homomorphisms of the DG algebra A(X) to the differential graded algebra of polynomial forms on a standard n-simplex, and identifying the faces with the homomorphisms given by restricting to the forms on a face of the standard n-simplex. The resulting complex is denoted $ <A(X)> $.
If we begin with a simplicial comples X, form its Sullivan algebra A(X), and then form the geometrization $ <A(X)> $; we get a space which is of the same homotopy type as X but is a thickening of it, with many parallel copies of X glued together. We shall discuss the possibility of constructing a state sum model analogous to the BC model on it below.

\bigskip

{\bf C. MODELS OF LOOP SPACES}

\bigskip
It is possible to assign to an object A of any of the above described model categories a loop space object $ \Omega (A)$ of the same category directly without reference to any geometric object.

The basic tool is the path fibration. If X is a connected topological space, then the path space with any two fixed endpoints is homotopy equivalent to the loop space of X. Now if we pick any point $x \in X$ then the set of all paths starting at x with arbitrary endpoint, which we denote P(x,X), is contractible, since the paths can be uniformly shrunk back to x. On the other hand, mapping each path to its endpoint gives us a map $ P(x,X) \rightarrow X $, whose fiber over any point is
$\Omega (X)$. It turns out that any fibration with contractible total space and base X is homotopy equivalent to this, so it is a homotopy path fibration, and its fiber is homotopic to $\Omega (X)$. 

We can imitate this in any model category by embedding any object A in a contractible space and then taking the quotient. This is one of the tricks which make model category theory so useful.

\bigskip

{\bf VI. SULLIVAN ALGEBRAS, MINIMAL AND CONTRACTIBLE MODELS}

\bigskip

Up to this point we have constructed various models for a space as differential graded algebraic objects. Whatever the intellectual attractions of these models, they are huge objects, and therefore not terribly helpful in computation. Sullivan discovered a procedure to replace these models with homotopy equivalent ones which are remarkably simple, and therefore enable us to perform all sorts of calculations.

Sullivan began by defining a special class of commutative DG algebras called Sullivan algebras.
\bigskip

{\bf DEFINITION:} A {\it Sullivan algebra} is an algebra of the form $( \Lambda V, d)$, where V is a graded vector space and $ \Lambda $ denotes the graded commutative free algebra, with the properties that V is the union of an upwards graded filtration $ V= \bigcup V_k $, such that $d:V_k \rightarrow \Lambda V_{k-1} $.

\bigskip

This definition implies that as a graded algebra, a Sullivan algebra is the tensor product of a symmetric tensor algebra, generated by the even degree elements of V, with an alternating or exterior algebra generated by the odd elements.

\bigskip

{\bf DEFINITION:} A Sullivan algebra is {\it minimal} if the image of d  is contained in $ \Lambda ^+ V \cdot \Lambda ^+ V $ where $ \Lambda ^+$ is the free algebra generated by elements of V with positive degree. In other words, all images of d are nontrivial products.

\bigskip

The definition of minimal excludes {\it contractible} Sullivan algebras.

\bigskip

{\bf DEFINITION:} A Sullivan algebra is {\it contractible} if has a basis of the form $ \{ x_i, y_i \} $ where  $d(x_i )= y_i,  d(y_i ) =0 $. 

\bigskip

It is clear from the definition that a contractible Sullivan algebra has no cohomology. 
\bigskip

{\bf THEOREM:} Every Sullivan algebra is the product of a minimal Sullivan algebra and a contractible one.

\bigskip

Now Sullivan goes on to show that every differential graded cochain algebra (this means a graded algebra of only nonnegative degree, where d goes up one in degree like differential forms) is quasiisomorphic to a Sullivan model, and in fact to a minimal one. This proves that simply connected rational homotopy types of spaces correspond one to one with isomorphism classes of minimal Sullivan models, since quasiisomorphic minimal models are in fact isomorphic.

His procedure is an inductive one reminiscent of the construction of Eilenberg-Maclane spaces.

We first put in generators for the cohomology of our model, then add in new generators to kill any unwanted cohomology which has appeared. We then repeat the second step to remove any unwanted cohomology it has generated, and repeat, infinitely if necessary.

In practice, this procedure terminates on quite simple DG algebraic structures.

We can then perform the loop space construction above on the ``minimal Sullivan Models, `` and obtain very simple models for loop spaces.

Let us describe the examples of the spheres.
\bigskip

{\bf Case 1. odd dimensional spheres}

\bigskip

The odd dimensional sphere $S^{2k-1} $ has a cohomology generator in d=2k-1.
This means it has for Sullivan minimal model the graded algebra $ \Lambda (e) $ with d=0.

Since e has odd degree, $e^2 =0$.

We can construct an algebraic path fibration for it by embedding it in the contractible differential graded algebra
$ \Lambda <e,u>, du=e$, where u has degree 2k-2.

Now to construct its algebraic loop space we pass to the quotient algebra

$ \Lambda <u>,  du=0.$

Since u has even degree, we see that the rational cohomology of $ \Omega (S^{2k-1}) $

has precisely one generator in each dimension which is a multiple of 2k-2.

\bigskip

{\bf Case 2, even dimension}

\bigskip

This case of $S^{2k}$ is slightly more complicated. 

We again start with $\Lambda <e>, d=0$,

However, we get an infinite dimensional algebra, since e now has even grade. According to Sullivan's construction, we need to kill off the higher cohomology by making it into coboundaries by adding in new generators. Fortunately, when we add in $e'$ with $ de' =e^2 $ with $e'$ of degree 4k-1, we find by a calculation that all the higher cycles are already killed off by monomials in e and $e'$.

Thus the Sullivan minimal model is given by $ \Lambda <e,e'> : de'=e$. We now proceed to construct its algebraic path fibration, given by $\Lambda <e, e', u,u'> de'=e^2, du=e,  du'=e'$. 

Once again we obtain the loop space by passing to the quotient $\Lambda <u,u'>$. We thus obtain, in particular the infinite set of images of the Schwartzschild black hole for k=1.

\bigskip

{\bf B. FREE LIE ALGEBRA MODELS}

\bigskip

The subclass of differential graded Lie algebra models analogous to the Sullivan models is the free Lie algebra models.
\bigskip

{\bf DEFINITION:} A free differential graded Lie algebra is the minimal Lie subalgebra of the free tensor algebra of a differential graded vector space V which contains V. 

\bigskip

In other words, it is the DG Lie algebra generated by all multiple brackets of elements of V with no relations except the axioms of a graded Lie algebra, namely graded Jacobi identitiy and graded commutativity.

\bigskip

{\bf Theorem:} Every topological space has a free DG Lie algebra model.

\bigskip

{\bf DEFINITION} A Free DG Lie algebra is minimal if the image under d of any element of V has projection 0 onto V.

\bigskip

Very similarly to the situation for Sullivan models, every free Lie algebra model is the product of a minimal one and a contractible one. Again, just as for Sullivan models, any two quasiisomorphic minimal free DG Lie algebras are isomorphic, so there is a 1-1 correspondence between minimal free DG Lie algebras and rational homotopy types of spaces. 

\bigskip

{\bf C. THE REDUCED BAR CONSTRUCTION}

\bigskip

As we have indicated, Quillen gave functors which established homotopy equivalences between his various model categories. The functor from DG Lie algebra models to DG algebra models is especially important for us. It is called the reduced bar construction \cite{Book}.

The reduced bar construction has the property that if applied to a free DG Lie algebra model which is finitely generated, it produces a Sullivan model. Since physically interesting models will be finitely generated because of energy and information bounds, this connection will be very important for us.

Let us briefly describe the reduced bar construction. Beginning with a free DG Lie algebra $(L_V, d_1) $,
we first form the dual graded vector space ${L_V}^*$. We then shift its grading up by 1, and form the free graded commutative tensor algebra on it, denoting the result as 

\bigskip

$ \Lambda (T (s{L_V}^*) $.

\bigskip

Note that the multiplication here is unrelated to the bracket on $L_V$. It is just a free product.

On this new algebra, graded by length of strings, we put a new differential $d= d_1 + d_2 $, where $d_1$ is the dual of the differential on $L_V$ and $d_2$ is the dual of the bracket on $L_V$. The bracket goes from $L_V \otimes L_V $ to $L_V$, so its dual is quadratic on ${L_V}^* $.

From our point of view, the reduced bar construction is an extremely convenient piece of magic. It converts a handlebody decomposition of the causal path space into an algebraic model for the differential forms on spacetime itself. The shift in grading reflects the transition from path space to spacetime, while the duality reflects the transition from cycles to forms.

\bigskip

{\bf VII. QUANTUM GEOMETRY OF Q REGIONS}

\bigskip

{\bf A. QUANTUM TOPOLOGY}

\bigskip

We can see that there is a strong coincidence between the structure of Sullivan algebras and the apparent multiple images of a light source seen through a highly curved region R of spacetime. The minimal part corresponds to the topologically unavoidable images which correspond to the generators of the rational cohomology of R, the contractible part to the pairs corresponding to cancelling handles, which can be seen to join and disappear as the observation point shifts. Astrophysicists refer to these as folds in space.

Furthermore, we would expect any physical region to be represented by a finitely generated model. We would also regard the free Lie algebra model as primary, since it relates directly to the path space of the region, and want to derive our model of the region itself from it.

We therefore form the

\bigskip

{\bf CONJECTURE:} {\it Quantum regions of spacetime are described by Sullivan algebras produced from finitely generated free Lie algebra models by the reduced bar construction.}

\bigskip

Note that the proposed models for quantum regions are richer than rational homotopy types, since they include the contractible algebras as well  The structure we are proposing is a model for a handlebody decomposition of the path space of a region, not only its topology. We can consider models with only cancelling pairs of handles, only free generators, or combinations of both.

Since Sullivan algebras are free algebras, they may be constructed as products of simple graded algebras with one generator each, and minimal contractible algebras
$\Lambda <x,y>, dy=x$. We think of these two types of generators as quantum versions of black holes and ``galaxies'' , since a small weak gravitational source produces a fold in the sky \cite{Perlick}.

Mathematically, rational homotopy types are as simple as they are because they admit free algebras as models. Since free algebras are simple products of parts with one generator each, this kind of decomposition is inherent in their structure.

The d operation in the combined algebra could make topological links between the parts. It is related to the Whitehead product.

As we have repeatedly emphasized, it is interesting from our point of view that our category of rational homotopy types has so many different but homotopy equivalent formulations. 
We conjecture that they correspond to complementary descriptions of the quantum spacetime geometry. In particular, the geometric realizations of Sullivan models is very suggestive. Recall that it is like a thickening of a region, where the different layers are labelled by different homomorphisms of the Sullivan algebra into the algebra of polynomial forms. In the case of a finitely generated free Sullivan algebra, the homomorphisms are determined by images of the generators, This seems like a natural setting for descriptions in which a subregion within our region observes different generators, and information about the geometry of the region is thereby measured. Such an description seems to be complementary to observing an external light source, just as a localized state for a particle is complementary to a beam.

It is also possible to start from a free DG lie algebra model $L_V$, and directly produce a complex called the cellular model of the Lie algebra. This consists of one cell for each basis element of the DG vector space V which generates the Lie algebra L, with attachments determined by the d operation on L restricted to V. This is of the same rational homotopy type as the geometric realization of the Sullivan algebra associated to $L_V$ by the Quillen functor, but is much simpler.

Thus more than one avenue toward a state sum model for quantum gravity over a Quillen model seems possible.

\bigskip

{\bf B. QUANTUM GEOMETRY}

\bigskip

So far, we have been investigating a new type of topology which may be relevant in quantum gravity.
Now we begin the investigation of how a quantum theory of general relativity might be constructed over a Sullivan model. There are two natural approaches, which we believe to be complementary.

To recapitulate, since the geometrical version of spaces Quillen began with was combinatorial, namely simplicial complexes, one natural avenue is to put state sum models similar to the BC model on them. Thus we could take the geometrization of a minimal Sullivan model as a starting point, and put a categorical state sum on it. The original BC model came out of quantizing the geometry of a four dimensional simplicial complex by quantizing variables corresponding to the immersion of each simplex into Minkowski space \cite{BC}. We could attempt something similar, attaching geometric variables to represent quantum geometrical degrees of freedom. 

A paricularly interesting situation arises if we begin with a finitely generated free DG Lie algebra model, transform it into a Sullivan model by the reduced bar construction, and form its geometric realisation. The model contains forms giving the relational geometry between the generating ``black holes'' and ``galaxies'', together with higher forms coming from the multiple brackets in the Lie algebra which would express higher quantum geometric correlations between them. If a state sum can be regularized to be finite over the ``layers'' of the Sullivan realisation, as we conjectured above, this would be an rich picture. We think that this has a good chance of working, because much of the information in the realisation groups together into families which are the same for any labelling of the boundary of any subcomplex. If such a state sum turns out to be possible, it will be an interesting departure, since the data on the simplices would not be local geometrical information, but only forms that indicated where the simplex ``saw'' the handles corresponding to the generators of the original Lie algebra model.

Another approach might be to regard the minimal Hopf algebra model as an approximation to the loop space, and rewrite Einstein's equations as conditions on the variation of the holonomy along the path under small perturbations. We could then write a topological version of this on the handlebody decompositions. of $\Omega(R)$ which correspond to the Sullivan model. Alternatively, we could write the Einstein-Hilbert Lagrangian on loop space by formulating the Riemann curvature in terms of variations of holonomy on the loop space, then approximate the path integral by a categorical state sum.

\bigskip

{\bf C. MATTER FIELDS}

\bigskip

We can approach representation of matter fields over a Sullivan model via the replacement of the Hopf algebra models by their graded tensor categories of representations. A representation of the Hopf algebra model is an approximation to a representation of the loop group, which is a way of describing a connection (composition of paths would be represented by composition of holonomies).

In the constructions of TQFTs and models for quantum gravity, the role of topology was played by simplicial complexes, and the fields came from Hopf algebras, tensor categories, and various categorifications thereof. There is something very suggestive about the topology of spacetime itself being represented by a Hopf algebra or tensor category. Various classes of functors are available for building blocks for a theory.

There is also the possibility of investigating quantum deformations of the Hopf algebra models in the sense of quantum groups.
Perhaps the resulting restriction of the representation theory could shed some light on what matter fields would appear.
\bigskip

{\bf D. A CONJECTURE: CRYSTALS}

\bigskip

Since the Sullivan models can be constructed as products of two basic pieces, which we suggestively named above as black holes and galaxies, it would be a natural question to ask what a quantum theory of gravity would do on such an aggregate. Gravity being an attractive force, we would expect them to clump together into some sort of crystalline structure, black holes standing in for nuclei and galaxies for electrons. Could the spacetime continuum appear in the limit from such crystals? Would matter appear as phonons in our spacetime crystal?

\bigskip

{\bf E. OUTLOOK}

\bigskip

The critical thing for this program is to develop one or more of the approaches to constructing a quantum theory of gravity on our proposed models. We should consider only models that are effectively four dimensional, i.e. only constructed from 2 and 3-handles as generators for a Sullivan algebra. Since we have constructed complexes which mirror the complexity of data which an external observer can see, topological state sum models on them would be plausible candidates for a quantum theory of gravity. Preliminary investigation of the Sullivan realisation geometric suggests that large ``gauge'' degrees of freedom exist, so that a state sum model has a good chance of being finite.

This would base quantum gravity in a mathematically well defined version of relational geometry as we outlined in the introduction. The geometric realisation is an interesting candidate for the thickening of spacetime at the quantum level.

If this is so, or if one of the other approaches to constructing a quantum theory outlined above proves tractible, then the new models proposed here could begin to be applied to the interesting problems of quantum gravity, and compared to other approaches.

\bigskip

{\bf ACKNOWLEDGEMENTS:}

\bigskip

The author wishes to thank The Maths Department of Nottingham University for their hospitality while this paper was being written. He thanks David Yetter, John Barrett, and Catherine  Meusburger for useful conversations. Special thanks to Dan Christensen who caught several mistakes in an earlier draft. The work was supported by a grant and a minigrant from FQXi.

\end{document}